# First Path Component Power Based NLOS Mitigation in UWB Positioning System

Authors version

Marcin Kolakowski[1], Jozef Modelski[2]

[1,2] Institute of Radioelectronics and Multimedia Technology, Warsaw University of Technology, Warsaw, Poland, contact: marcin.kolakowski@pw.edu.pl

# First Path Component Power Based NLOS Mitigation in UWB Positioning System

Marcin Kolakowski, *Student Member, IEEE,* Jozef Modelski, *Fellow, IEEE*

*Abstract* — The paper describes an NLOS (Non Line of Sight) mitigation method intended for use in a UWB positioning system. In the proposed method propagation conditions between the localized objects and anchors forming system infrastructure are classified into one of three categories: LOS (Line-of-Sight), NLOS (Non-Line-of-Sight) and severe NLOS. Non-Line_of-Sight detection is conducted based first path signal component power measurements. For each of the categories average NLOS inducted time of arrival bias and bias standard deviation have been estimated based on results gathered during a measurement campaign conducted in a fully furnished apartment. To locate a tag an EKF (Extended Kalman Filter) based algorithm is used. The proposed method of NLOS mitigation consists in correcting measurement results obtained in NLOS conditions and lowering their significance in a tag position estimation process. The paper includes the description of the method and the results of the conducted experiments.

## I. INTRODUCTION

ULTRA-WIDBAND positioning systems allow for precise indoor localization with accuracy of several dozen centimeters. However, such accuracy, for most system implementations is attainable only in specific LOS (Line of Sight) working conditions, when the direct visibility between the localized object and system infrastructure is not obstructed by any obstacles.

Unfortunately in most indoor environments it is very difficult to provide a direct visibility between each point and all of the anchor nodes used in position calculation, because it would require using a very large number of devices and rise system's cost. Therefore most of the currently used systems operate in NLOS (Non-Line-of-Sight) conditions, in which a visibility between the localized objects and system infrastructure is partially or fully obstructed.

Positioning in most of the UWB systems relies on measurements of time of arrival of the signals sent by the localized tag or system infrastructure. In NLOS conditions the signal propagates through obstacles which may introduce a delay of a magnitude of several nanoseconds. It has a negative impact on localization accuracy and precision. Fortunately those negative effects can be partially reduced through using NLOS mitigation techniques.

The first step of most mitigation methods is NLOS conditions detection. It is done in many different ways based on various parameters. NLOS conditions can be detected through statistical analysis of the received signal parameters such as RSS (Received Signal Strength), TOA (Time of Arrival) and RMS (Root Mean Squared Delay Spread) [1]. Another approaches include the use of machine learning algorithms for analysis of channel impulse response parameters [2][3] and map analysis combined with ray tracing [4]. Although all of the referenced methods allow to identify NLOS with high success rate, they are not practical when it comes to real system implementation. Extraction and statistical processing of channel impulse response or advanced analysis of localized objects surroundings is computationally demanding and may not be possible in real time.

When detected, NLOS working conditions can be mitigated in different ways. The simplest technique is to exclude NLOS deteriorated results [5] . However such approach is not always applicable, because discarding NLOS results might lead to situation, when the number of measurement results is not sufficient to calculate any localization. Therefore many works describe methods, which allow to include NLOS results in position computations and maintain good positioning accuracy. Those methods consist in correcting NLOS results based on measurement results [2,3] or joint processing of LOS and NLOS results using linear programming [3], machine learning [6] and Kalman filtering [7] methods.

In the paper a NLOS detection and mitigation technique is presented. The first step of the proposed method consists in classifying propagation conditions between the tag and the anchors into one of three categories. According to the category, measurement results are corrected and taken with different weights during position calculation, which is performed using an EKF (Extended Kalman Filter) based algorithm. NLOS detection criteria and values of corrections are based on the analysis of the results of an measurement campaign conducted in a fully furnished apartment which is fully described in Section II of the paper. Section III contains the description of the EKF based localization algorithm and NLOS mitigation method used in the system. Section IV includes the results of conducted experiments. Section V concludes the paper.

All authors are with the Institute of Radioelectronics and Multimedia Technology, Warsaw University of Technology, Nowowiejska 15/19, 00-665 Warsaw, Poland (phone: +48 22 2347635; e-mails: m.kolakowski@ire.pw.edu.pl, j.modelski@ire.pw.edu.pl).

## II. NON-LINE-OF-SIGHT PROPAGATION INVESTIGATION

### A. Measurement site

The development and implementation of the proposed NLOS mitigation method have been preceded by the measurement campaign conducted in a typical apartment. The apartment was located in a panel building with walls made of pre-stressed concrete and was fully furnished. The layout of the measurement site is presented in Fig.1.

In the apartment a DW1000 [8] based UWB positioning system [9] was deployed. The system consisted of one tag and six anchors, which were attached to the walls in different rooms. The tag is located based on Time Difference of Arrival (TDOA) of the signals transmitted by the tag and received by the anchors. For the purpose of the experiment the functionality of propagation time measurement between the tag and the anchors has been added. Propagation time is measured using Symmetric Double Sided Two Way Ranging method (SDS-TWR) [10].

The test consisted in estimating the propagation time between the tag and the anchors and measuring received tag signal power. During the experiment the tag has been placed in 26 test points distributed in the apartment.

### B. First Path Power measurements

DW1000 UWB radio chips allow to measure first path component power. Exemplary averaged power measurement results for anchor AN 6 are presented in Fig.2. The power of the first path signal component received by the anchor depends on propagation path. In the picture, it can be seen that received power levels tend to take certain values grouped in three different ranges.

For each test point an analysis of propagation conditions was performed. It resulted in assigning test point–anchor radio links to following propagation conditions categories:
- Line-of-Sight (LOS) conditions, where there are no obstacles between the tag and the anchor,
- Non-Line-of-Sight (NLOS) conditions, e.g. propagation through one wall or a single obstacle,
- Severe Non-Line-of-Sight (SNLOS) conditions, e.g. propagation through multiple walls and obstacles

Category of propagation conditions between the tag and the specific anchor can be determined by comparing received power measurement results with two thresholds. Based on the measurements performed for all of the twenty-six test points threshold values allowing to classify propagation conditions with the highest success rate have been chosen. The obtained thresholds have the following values: if the power received by the anchor is higher than -78.5 dBm the method assumes that the tag is directly visible. If the received power is lower but still higher than -85 dBm the tag and the anchor work in NLOS conditions. For lower power values the conditions are classified as SNLOS.

The proposed method was tested on another set of measurement data collected during one of the earlier system tests and the results were verified with test point layout The proposed method correctly assigned to particular categories about 78% of the radio links between the tag and the anchors.

### C. Propagation time measurements

Propagation time between the tag located in each of the test points and all anchors has been measured with SDS-TWR method. The bias present in NLOS conditions was calculated by subtracting the propagation time calculated based on geometry of the system (Euclidean distance between the test points and the anchors divided by speed of light) from measured propagation time. Histograms of the obtained bias values for NLOS and SNLOS conditions are presented in Fig.3. Average values and standard deviations of bias inducted by NLOS and SNLOS conditions are stored in Table.1.

Some of the obtained bias values are negative. It results from one of the properties of employed UWB modules. In DW1000 based devices a received signal level dependent bias in measured distance can be observed. The corrections suggested in [11] depend on signal level calculated for all components reaching the receiver in free-space propagation conditions. Signal attenuation in NLOS conditions is not considered. Therefore no corrections were introduced to measurement results neither during first path propagation time evaluation nor measurements used for positioning. It. results in negative bias in range measurements.

The obtained bias average values and standard deviations are used by NLOS mitigation method to improve positioning accuracy.

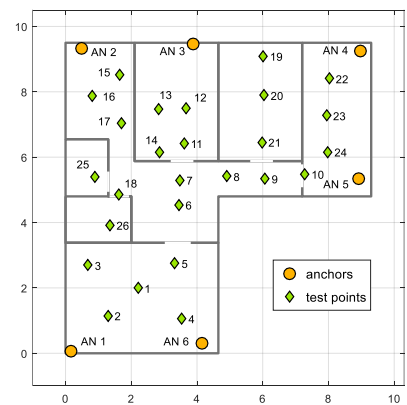

Fig. 1. Measurement site with marked UWB system anchors and test points location

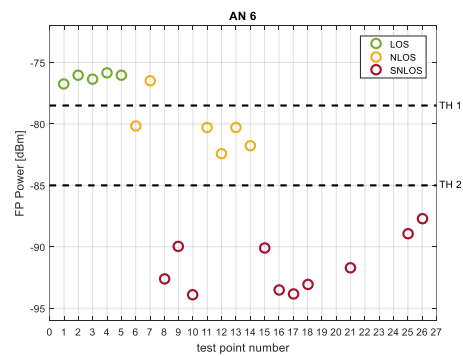

Fig. 2. Received tag signal power measured by anchor AN 6 for the tag located in test points

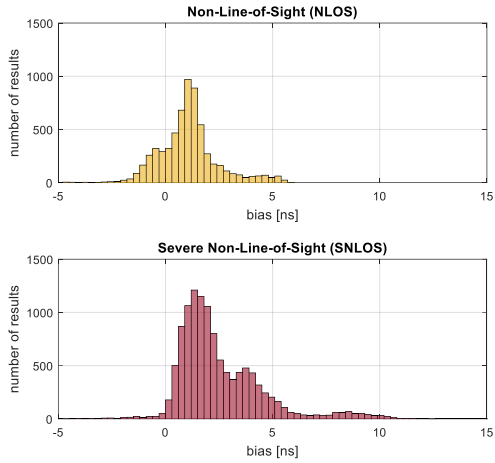

Fig. 3. Bias histograms for NLOS and SNLOS conditions

TABLE 1: NLOS BIAS MEAN VALUE AND STANDARD DEVIATION

| Category | mean | std |
|---|---|---|
| Non-Line-of-Sight | 0.49 ns | 1.39 ns |
| Severe Non-Line-of-Sight | 1.92 ns | 2.02 ns |

## III. LOCALIZATION ALGORITHM AND NLOS MITIGATION

### A. Localization algorithm

The tag is localized with an Extended Kalman Filter [12] based algorithm. The EKF is described with the following set of equations:

$$x_{k(-)} = F x_{k-1(+)} \quad (1)$$

$$P_{k(-)} = F P_{k-1(+)} F^T + Q_k \quad (2)$$

$$x_{k(+)} = x_{k(-)} + K_k (z_k - h_k(x_{k(-)})) \quad (3)$$

$$K_k = P_{k(-)} H_k^T [H_k P_{k(-)} H_k^T + R_k]^{-1} \quad (4)$$

$$P_{k(+)} = [I - K H_k] P_{k(-)} \quad (5)$$

In the algorithm the tag is treated as a dynamic system described, which at a given moment $k$ is described with a state vector $x_k$ containing tag's location coordinates and velocity components alongside x and y axes. At the start of the algorithm it is assumed that tag's location is the middle of the apartment and the tag is not moving.

Equations (1-2) comprise time-update phase, during which a current value of state vector $x_{k(-)}$ is predicted based on a value obtained in the previous iteration $x_{k-1(+)}$ and equations of motion stored in matrix $F$. Matrices $P_{k(-)}$ and $P_{k-1(+)}$ are covariance matrices corresponding to those vectors and $Q_k$ is process noise covariance matrix chosen with accordance to DWNA (Discrete White Noise Acceleration) model [13].

Equations (3-5) describe measurement-update phase, during which the predicted state vector is updated with measurement results. The resulting state vector is a linear combination of the predicted value and measurement innovation which is a difference between TDOA measurement results obtained with the anchors stored in vector $z_k$ and predicted measurements $h_k(x_{k(-)})$ estimated for prognosed tag location. The extent to which measurement results are taken into consideration while updating the state vector is determined by Kalman gain matrix $K_k$. It is calculated based on covariance matrix for predicted vector, linearized measurement function and measurement covariance matrix $R_k$. The updated state vector value is the final result of the algorithm and is used in the next iterations.

### B. NLOS mitigation

The proposed method of NLOS mitigation consists in correcting measurement results obtained in NLOS conditions and lowering their significance in tag position estimation process.

Measurement vector $z$ contains TDOA values calculated based on time of arrival (TOA) measurements conducted by the anchors. $TDOA_{ij}$ between anchors $AN_i$ and $AN_j$ is calculated by subtracting time of arrival measured by the second one $TOA_j$ from the first one $TOA_i$.

$$TDOA_{ij} = TOA_i - TOA_j \quad (6)$$

When the system works in NLOS conditions, the signal sent by the tag is delayed. NLOS inducted delay can be modeled as a random variable which is added to the propagation time resulting from geometrical tag and anchor locations. Therefore, the measured value of UWB packet's time of arrival can be expressed as:

$$TOA_m = TOA + n_{TOA} + b + n_b \quad (7)$$

Where $TOA$ is signal's time of arrival, $n_{TOA}$ is time of arrival measurement noise, b is average NLOS inducted bias and $n_b$ is bias random component. Both $n_{TOA}$ and $n_b$ are modeled to be zero-mean variables with variance $\sigma^2_t$ and $\sigma^2_b$ respectively.

The workflow of the propose NLOS mitigation method is presented in Fig.4. The method starts with classifying propagation conditions between the tag and all of the anchors to one of three categories based on measured first path power. The threshold values were chosen based on the measurements described in Section II. Then the TOAs measured by the obstructed anchors are corrected by subtracting bias induced by NLOS conditions:

$$TOA_{m\_corr} = TOA_m - b = TOA + n_{TOA} + n_b \quad (8)$$

The values of the subtracted bias depends on into which category propagation conditions have been classified and are average bias values, which are stored in Table 1.

Bias random component which is still present in measured TOAs cannot be easily removed. Therefore it is taken into account in another way by treating corrected TOA as a variable with a higher variance of:

$$\sigma^2_{m\_corr} = \sigma^2_t + \sigma^2_b \quad (9)$$

The assumed value of time measurement standard deviation $\sigma_t$ is 0.2 ns. Bias variances $\sigma^2_b$ for each category are calculated for standard deviations stored in Table 1. The higher variance of corrected TOAs are taken into account by adjusting corresponding TDOAs covariances in measurement covariance matrix R.

The obtained corrected TOA measurement results and adjusted R matrix are then processed with the EKF based localization algorithm.

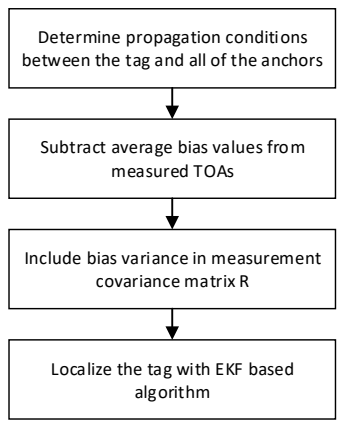

Fig. 4. Workflow of the proposed NLOS mitigation method

## IV. EXPERIMENTS

The proposed algorithm and NLOS mitigation method have been tested with an experiment. The performed tests consisted in localization of the static tag placed in the same environment and test points as in the measurement campaign described in Section II. The layout of system anchors and the locations of the test points are presented in Fig.1.

In the experiment tag's location was calculated using two algorithms: EKF without NLOS influence mitigation and EKF with addition of NLOS mitigation method proposed in the paper. Exemplary localization results obtained for selected points are presented in Fig.5. The empirical cumulative distribution functions for localization errors are presented in Fig.6.

The use of the proposed NLOS mitigation technique allowed to localize the tag with higher accuracy in sixty percent of test points. Forty percent of points were located with the same accuracy as without NLOS mitigation. It results from the fact that for some test points, anchors for which TDOAs were calculated were classified to the same propagation conditions, which led to same bias corrections being subtracted from each other and TDOAs remaining unchanged.

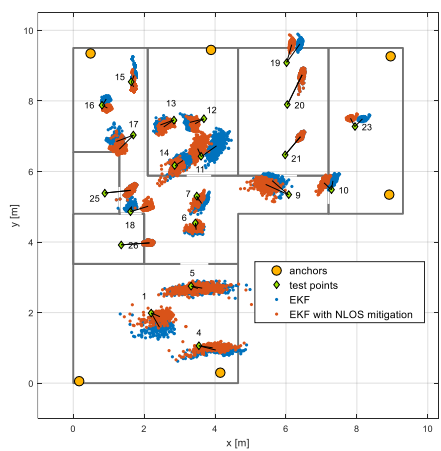

Fig. 5. Exemplary localization results obtained with both algorithms

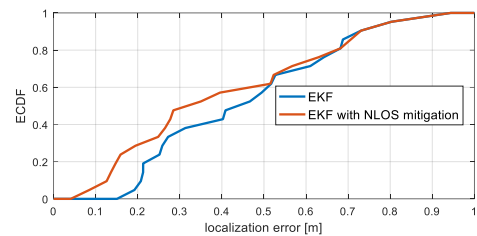

Fig. 6. Empirical CDF curves of localization errors obtained for algorithms without and with error mitigation.

## V. CONCLUSION

In the paper a simple NLOS mitigation method intended for use in UWB positioning system is presented. The method classifies propagation conditions between the tag and all of the anchors and introduces appropriate corrections to measurement results returned by the system infrastructure. Correction values were obtained based on a measurement campaign conducted in a typical indoor environment.

The conducted experiments have shown that in most cases the proposed method allows to reduce positioning error. The efficiency of the method can be increased through using more robust classification of the NLOS conditions and implementing more advanced detection techniques, which will be the subject of future authors' research.